\documentclass[preprint,amsmath,amssymb,nofootinbib]{revtex4}

\usepackage{graphicx}
\usepackage{dcolumn}
\usepackage{bm}

\begin{document}

\title{The Rise of Solitons in Sine-Gordon Field Theory: From Jacobi Amplitude to Gudermannian Function}

\author{Leonardo Mondaini$^{1,2}$}
\email{mondaini@ualberta.ca, mondaini@unirio.br}
\affiliation{$^1$Department of Oncology, University of Alberta, Edmonton, Canada \\$^2$Grupo de F\'{i}sica Te\'orica e Experimental, Departamento de Ci\^encias Naturais, Universidade Federal do Estado do Rio de Janeiro, Rio de Janeiro, Brazil}


\begin{abstract}
We show how the famous soliton solution of the classical sine-Gordon field theory in $(1+1)$-dimensions may be obtained as a particular case of a solution expressed in terms of the Jacobi amplitude, which is the inverse function of the incomplete elliptic integral of the first kind. 

{\bf{Keywords:}} Solitons, Sine-Gordon Field Theory, Elliptic Integrals, Jacobi Amplitude

\end{abstract}


\maketitle


\section{Introduction}
The sine-Gordon field theory and the associated massive Thirring model \cite{coleman} are some of the best studied quantum field theories. In view of its connections to other important physical models, some of which in principle
admit actual realizations in nature \cite{kosterlitz, samuel}, a huge mass of important exact results have been obtained for this fascinating integrable system \cite{solitons, MondainiJSP, MondainiJPA, MondainiJMP}. However, no less fascinating are the remarkable mathematical and physical properties of its soliton (or ``solitary wave") solutions which have contributed, along the last decades, to turn the physics of solitons into a very active research topic.

In this work we present a simple and yet appealing step-by-step derivation of a more general solution for the classical sine-Gordon field theory in (1+1)-dimensions in terms of a special kind of elliptic function, namely the Jacobi amplitude, which has the famous sine-Gordon soliton solution as a particular case. Despite the fact that the connection between solitons and Jacobi elliptic functions has already been explored in \cite{cervero}, we believe this work comes to shed more light on this interesting subject, helping to fill in a gap existing in the corresponding specialized literature.

\section{An alternative pathway to solitons in sine-Gordon field theory}

\subsection{The Jacobi amplitude function}

We start by considering the following theory describing a real
scalar field in (1+1)-dimensions ($\phi\equiv\phi(x,t)$),
\begin{equation}
\mathcal{L}=\frac{1}{2}\partial_\mu\phi\partial^\mu\phi-V(\phi), \label{eq1}
\end{equation}
where the potential term is given by
\begin{equation}
V(\phi) = 2\alpha\cos(\beta\phi) + 2\gamma. \label{eq2}
\end{equation}
The above Lagrangian gives rise, through the  Euler-Lagrange equation, $\partial_\mu\left(\frac{\partial \mathcal{L}}{\partial(\partial_\mu\phi)}\right)=\frac{\partial\mathcal{L}}{\partial\phi}$, to the following field equation
\begin{equation}
\partial_\mu\partial^\mu \phi \equiv \Box\phi \equiv \left(\frac{1}{c^2}\frac{\partial^2}{\partial t^2}-\frac{\partial^2}{\partial x^2}\right)\phi= -\frac{\partial V(\phi)}{\partial\phi}.
\label{eq3}
\end{equation}
Notice that since Equation (\ref{eq3}) is invariant under Lorentz transformations ($x^\mu\rightarrow x'^\mu=\Lambda^\mu\,_\nu x^\nu$)\cite{rbef}, its solutions may be obtained through the solutions of the
corresponding equation for the static case ($\phi\equiv \phi(x)$) by a simple Lorentz boost, namely $x-x_0\rightarrow (x-x_0-vt)/\sqrt{1-(v^2/c^2)}$, for arbitrary $v$ ($|v|<c\approx 3\times10^8\, {\rm m/s}$)\cite{jackiw, rajaraman}. Thus, in what follows, we will focus on the solutions of the equation
\begin{equation}
\frac{d^2\phi}{d x^2}=\frac{d V}{d\phi}.
\label{eq4}
\end{equation}
Indeed, by multiplying the above equation by $d\phi/dx$ we obtain
\begin{equation}
\frac{d\phi}{d x}\frac{d^2\phi}{d x^2}=\frac{d\phi}{d x}\frac{d V}{d\phi}\Rightarrow\frac{d}{d x}\left[\frac{1}{2}\left(\frac{d\phi}{dx}\right)^2\right] = \frac{d V}{dx},
\label{eq5}
\end{equation}
which, after an integration with respect to $x$ and some algebra, may be rewritten as
\begin{equation}
d x'=\pm\frac{d\phi'}{\sqrt{2V(\phi')}}.
\label{eq6}
\end{equation}
By integrating both sides of the above equation, from $x'=x_0$ to $x'=x$ ($\phi'=\phi(x_0)$ to $\phi'=\phi(x)$), we get
\begin{equation}
x-x_0 = \pm\int_{\phi(x_0)}^{\phi(x)}\frac{d\phi'}{\sqrt{2V(\phi')}}.
\label{eq7}
\end{equation}
In order to compute the above integral, we must firstly notice that the potential, shown in Equation (\ref{eq2}), may be rewritten as 
\begin{equation}
V(\phi') = 2(\alpha+\gamma)\left[1-\frac{2\alpha}{\alpha+\gamma}\sin^2\left(\frac{\beta\phi'}{2}\right)\right].
\label{eq8}
\end{equation}
Thus, by making the change of variables $\phi'\rightarrow \theta' = \frac{\beta}{2}\phi'$, defining $k^2 = \frac{2\alpha}{\alpha +\gamma}$ and choosing $x_0$ such that $\phi(x_0)=0\Rightarrow \theta_0 =0$, we are left with
\begin{equation}
x-x_0 = \pm\frac{k}{\beta\sqrt{2\alpha}}\int_{0}^{\theta}\frac{d\theta'}{\sqrt{1-k^2\sin^2{\theta'}}}.
\label{eq9}
\end{equation}
The integral appearing in Equation (\ref{eq9}) is called an incomplete elliptic integral of the first kind, $F(\theta,k)$, whereas $k$ is called the elliptic modulus or eccentricity. The upper limit, $\theta$, of this integral may be written in terms of the {\it Jacobi amplitude} (the inverse function of the incomplete elliptic integral of the first kind) as \cite{grad, jacobi}.
\begin{equation}
\theta=\pm\,F^{-1}\left(\frac{\beta\sqrt{2\alpha}}{k}(x-x_0), k\right)\equiv \pm\,{\rm am}\left(\frac{\beta\sqrt{2\alpha}}{k}(x-x_0), k\right).
\label{eq10}
\end{equation}
Notice that, from the above definition, we have $F({\rm am}\left(x, k\right),k)=x$. 

The solution of Equation (\ref{eq4}) may be, finally, written as
\begin{equation}
\phi(x)= \pm\frac{2}{\beta}\,{\rm am}\left(\frac{\beta\sqrt{2\alpha}}{k}(x-x_0), k\right).
\label{eq11}
\end{equation}
Hence, from the above equation, we may notice that
\begin{equation}
\phi(x_0)= \pm\frac{2}{\beta}\,{\rm am}\left(0, k\right)=0,
\label{eq12}
\end{equation}
as it should.

\subsection{The case $k=1$: The Gudermannian function  and the soliton solution of sine-Gordon equation}

From the definition $k^2 = 2\alpha/(\alpha +\gamma)$ we may obviously see that when $\gamma = \alpha$ we have $k = 1$. Hence, the solution for Equation (\ref{eq4}) with the potential given by
\begin{equation}
V(\phi) = 2\alpha[1+\cos(\beta\phi)], \label{eq13}
\end{equation}
may be obtained as a special case of the solution presented in Equation (\ref{eq11}). Indeed, since ${\rm am}(x, 1) = {\rm gd}\,x \equiv 2 \arctan(e^x) - \pi/2$, where ${\rm gd}\,x$ is called the {\it Gudermannian function} (a special function which relates the circular functions to the hyperbolic ones without using complex numbers, named after Christoph Gudermann (1798 - 1852)), we are left with
\begin{eqnarray}
\phi(x)= \pm\frac{2}{\beta}\,{\rm am}\left(\beta\sqrt{2\alpha}(x-x_0), 1\right) &=&\pm\frac{2}{\beta}\,{\rm gd}\left(\beta\sqrt{2\alpha}(x-x_0)\right)\nonumber\\ &\equiv& \pm\frac{4}{\beta}\arctan\left[\exp \left(\beta\sqrt{2\alpha}(x-x_0)\right)\right] \mp \frac{\pi}{\beta}.
\label{eq14}
\end{eqnarray}
Last but not least, we must notice that by substituting the Equation (\ref{eq14}) into Equation (\ref{eq3}) and making the change (Lorentz boost) $x-x_0\rightarrow (x-x_0-vt)/\sqrt{1-(v^2/c^2)}$, we obtain the famous sine-Gordon field equation, namely
\begin{equation}
\Box\phi_S +2\alpha\beta\sin \beta\phi_S=0, 
\label{eq15}
\end{equation}
where $\phi_S\equiv\phi_S(x,t)$ is the no less famous soliton/anti-soliton solution \cite{jackiw, rajaraman}, given by
\begin{equation}
\phi_S(x,t)=\pm\frac{4}{\beta}\arctan\left[\exp \left(\beta\sqrt{2\alpha}\frac{x-x_0-vt}{\sqrt{1-(v^2/c^2)}}\right)\right] .
\label{eq16}
\end{equation}
This result allows us to characterize the Lorentz boosted, and shifted by $\pi/\beta$, version of the solution in terms of the Jacobi amplitude shown in Equation (\ref{eq11}), namely   
\begin{equation}
\phi(x,t)= \pm\frac{2}{\beta}\,{\rm am}\left(\frac{\beta\sqrt{2\alpha}}{k}\frac{x-x_0-vt}{\sqrt{1-(v^2/c^2)}}, k\right) \pm \frac{\pi}{\beta},
\label{eq17}
\end{equation}
as a generalization of the sine-Gordon soliton/anti-soliton solution for $k \neq 1$.

\section{Concluding remarks}
\begin{figure} [h]
\begin{center}
\hfil\scalebox{0.5}{\includegraphics{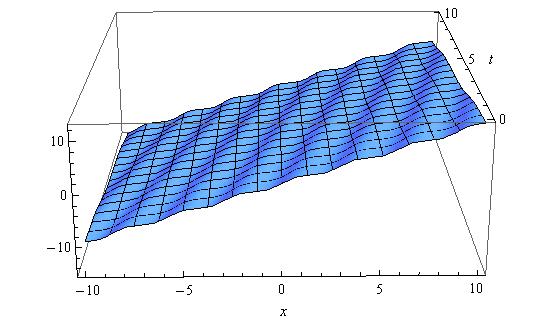}}\hfil
\caption{The Jacobi amplitude solution given by Equation (\ref{eq17}) with $\alpha=0.50$, $\beta=2.00$, $x_0=0$, $k=0.99$ and $v=0.50 c$}
\label{Figure1}
\end{center}
\end{figure}

\begin{figure} [h]
\centering
\hfil\scalebox{0.5}{\includegraphics{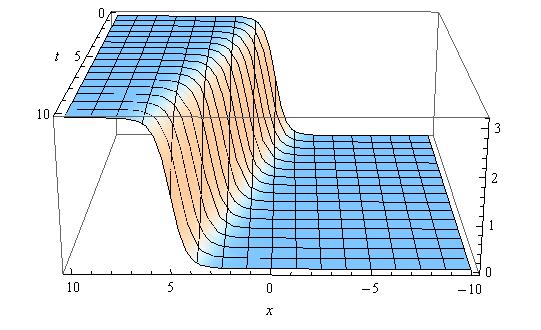}}\hfil
\caption{The soliton solution given by Equation (\ref{eq16}) with $\alpha=0.50$, $\beta=2.00$,  $x_0=0$, $k=1.00$ and $v=0.50 c$}
\label{Figure2}
\end{figure}

We would like to make a few comments about the soliton solution, shown in Equation (\ref{eq16}), and its generalized version, shown in Equation (\ref{eq17}). 
Firstly, we may notice by comparing the Figures \ref{Figure1} and \ref{Figure2} how different are these solutions, where we would like to highlight the doubly periodic behaviour of the Jacobi amplitude solution.

Finally, let us observe that, as remarked in \cite{jackiw}, this soliton solution, though arising in a classical field theory,
looks very much like a classical particle since its energy density is localized at a point ($x=x_0$) and its total energy for a static field configuration ($\phi_S\equiv\phi_S(x)$), namely 
\begin{equation}
E(\phi_S)=\int_{-\infty}^{\infty} dx\left[\frac{1}{2}\left(\frac{d\phi}{dx}\right)^2+V(\phi)\right]=\frac{8\sqrt{2\alpha}}{\beta},
\label{eq18}
\end{equation}
is finite, just as we should expect.

\section*{Acknowledgements}

This work has been supported by University of Alberta's Li Ka Shing Applied Virology Institute and CNPq, Conselho Nacional de Desenvolvimento Cient\'{i}fico e Tecnol\'ogico - Brasil.


%
%

\begin{thebibliography}{99}
%
%
\bibitem{coleman}
Coleman, S. (1975)
{\it Quantum Sine-Gordon Equation as the Massive Thirring Model}. Phys. Rev. D, 11, 2088-2097. http://dx.doi.org/10.1103/PhysRevD.11.2088 
%
%
\bibitem{kosterlitz}
Kosterlitz, J.M. (1974)
{\it The Critical Properties of the Two-Dimensional XY Model}. J. Phys. C: Solid State Phys., 7, 1046-1060. http://dx.doi.org/10.1088/0022-3719/7/6/005
%
%
\bibitem{samuel}
Samuel, S. (1978)
{\it Grand Partition Function in Field Theory
with Applications to Sine-Gordon Field Theory}. Phys. Rev. D, 18, 1916-1932. http://dx.doi.org/10.1103/PhysRevD.18.1916 
%
%
\bibitem{solitons}
Dauxois, T. and Peyrard, M. (2006) {\it Physics of Solitons}. Cambridge University Press, New York.
%
%
\bibitem{MondainiJSP}
Mondaini, L. and Marino, E.C. (2005)
{\it Sine-Gordon/Coulomb
Gas Soliton Correlation Functions and an Exact Evaluation
of the Kosterlitz-Thouless Critical Exponent}. J. Stat. Phys., 118, 767-779. http://dx.doi.org/10.1007/s10955-004-8828-y
%
%
\bibitem{MondainiJPA}
Mondaini, L., Marino, E.C. and Schmidt, A.A. (2009)
{\it Vanishing Conductivity of Quantum Solitons in Polyacetylene}. J. Phys. A: Math. Theor., 42, 055401. http://dx.doi.org/10.1088/1751-8113/42/5/055401
%
%
\bibitem{MondainiJMP}
Mondaini, L. (2012)
{\it Thermal Soliton Correlation Functions in Theories
with a $Z(N)$ Symmetry}. J. Mod. Phys., 3, 1776-1780. http://dx.doi.org/10.4236/jmp.2012.311221
%
%
\bibitem{cervero}
Cervero, J.M. (1986)
{\it Unveiling the Solitons Mistery: The Jacobi Elliptic Functions}. Am. J. Phys., 54, 35-38. http://dx.doi.org/10.1119/1.14767 
%
%
\bibitem{rbef}
Mondaini, L. (2012)
{\it Obtaining a Closed-form Representation for the Dual Bosonic Thermal Green Function by Using Methods of Integration on the Complex Plane}. Rev. Bras. Ens. Fis., 34, 3305. http://dx.doi.org/10.1590/S1806-11172012000300005
%
%
\bibitem{jackiw}
Jackiw, R. (1977) {\it Quantum Meaning of Classical Field Theory}. Rev. Mod. Phys., 49, 681-706. http://dx.doi.org/10.1103/RevModPhys.49.681 
%
%
\bibitem{rajaraman}
Rajaraman, R. (1987)
{\it Solitons and Instantons: An Introduction to Solitons and Instantons in Quantum Field Theory}. Elsevier, Amsterdam.
%
%
\bibitem{grad}
Gradshteyn, I.S. and Ryzhik, I.M. (2000) {\it Table of Integrals, Series, and
Products}. Academic Press, San Diego.
%
%
\bibitem{jacobi}
Weisstein, E.W. {\it Jacobi Amplitude}. MathWorld -- A Wolfram Web Resource. http://mathworld.wolfram.com/JacobiAmplitude.html
%
%
\end{thebibliography}
\end{document}